\documentclass[aps, prl, twocolumn, floatfix, superscriptaddress]{revtex4-2}
\usepackage{graphicx, amsmath, amssymb, color}
\usepackage{multirow, makecell}
\usepackage{hyperref}  
\hypersetup{breaklinks={true}}

\begin{document}
\title{Higher-order components dictate higher-order contagion dynamics in hypergraphs}
\author{Jung-Ho Kim}
\affiliation{Department of Physics, Korea University, Seoul 02841, Korea}
\author{K.-I. Goh}
\email{kgoh@korea.ac.kr}
\affiliation{Department of Physics, Korea University, Seoul 02841, Korea}
\affiliation{Department of Mathematics, University of California Los Angeles, Los Angeles, CA 90095, USA}
\date{\today}
\begin{abstract}
The presence of the giant component is a necessary condition for the emergence of collective behavior in complex networked systems.
Unlike networks, hypergraphs have an important native feature that components of hypergraphs might be of higher order, which could be defined in terms of the number of common nodes shared between hyperedges.
Although the extensive higher-order component (HOC) could be witnessed ubiquitously in real-world hypergraphs, the role of the giant HOC in collective behavior on hypergraphs has yet to be elucidated.
In this Letter, we demonstrate that the presence of the giant HOC fundamentally alters the outbreak patterns of higher-order contagion dynamics on real-world hypergraphs. 
Most crucially, the giant HOC is required for the higher-order contagion to invade globally from a single seed.
We confirm it by using synthetic random hypergraphs containing adjustable and analytically calculable giant HOC.
\end{abstract}
\maketitle

\textit{Introduction.}---The complex system is composed of many interacting elements, and the interaction might occur not only between two elements but generally within a group of an arbitrary number of elements simultaneously~\cite{2020BattistonNetworks, 2021BattistonThe, 2022MajhiDynamics}.
Group interaction is essential for understanding various complex systems' functions, such as social contact~\cite{2018BensonSimplicial}, coauthorship~\cite{2010TaramascoAcademic, 2017PataniaThe, 2018BensonSimplicial}, brain~\cite{2016GiustiTwo}, biology~\cite{2009KlamtHypergraphs, 2021FengHypergraph}, and ecology~\cite{2016BaireyHigh-order, 2017GrilliHigher-order, 2017LevineBeyond}, to name a few.
Notably, the structures and dynamics in these systems cannot be fully understood by using the projected network with pairwise interactions.
From this perspective, studies that introduce group interactions into classical statistical physics problems like percolation~\cite{2018BianconiTopological, 2020CoutinhoCovering, 2021LeeHomological, 2021SunHigher-order, 2022ZhaoHigher-order}, random walk~\cite{2020CarlettiRandom, 2021CarlettiRandom}, contagion dynamics~\cite{2019IacopiniSimplicial, 2019JhunSimplicial, 2020ArrudaSocial, 2020LandryThe, 2021St-OngeUniversal, 2021St-OngeSocial}, synchronization~\cite{2019SkardalAbrupt, 2020MillanExplosive, 2021GambuzzaStability, 2021KovalenkoContrarians}, opinion dynamics~\cite{2020HorstmeyerAdaptive, 2021NoonanDynamics, 2022PapanikolaouConsensus}, evolutionary game theory~\cite{2021Alvarez-RodriguezEvolutionary, 2021CiviliniEvolutionary, 2021GuoEvolutionary}, and statistical validated hypergraphs \cite{2021MusciottoDetecting} have been actively conducted recently.
These studies have revealed that higher-order interactions significantly alter collective dynamics.

A hypergraph is a data structure expressing group interactions and consists of nodes representing elements and hyperedges representing interactions between elements~\cite{1989BergeHypergraphs}.
In a network, an edge represents the interaction only between two nodes, whereas, in a hypergraph, the hyperedge represents the interaction between an arbitrary number of nodes.
In other words, a hypergraph is a generalization of a network.
We use the term the degree $k$ of a node for the number of hyperedges that the node belongs to; and the size $s$ of a hyperedge for the number of nodes belonging to the hyperedge.

\begin{figure}[b]
\centering
\vspace*{-0.2cm}
\includegraphics[width=0.48\textwidth]{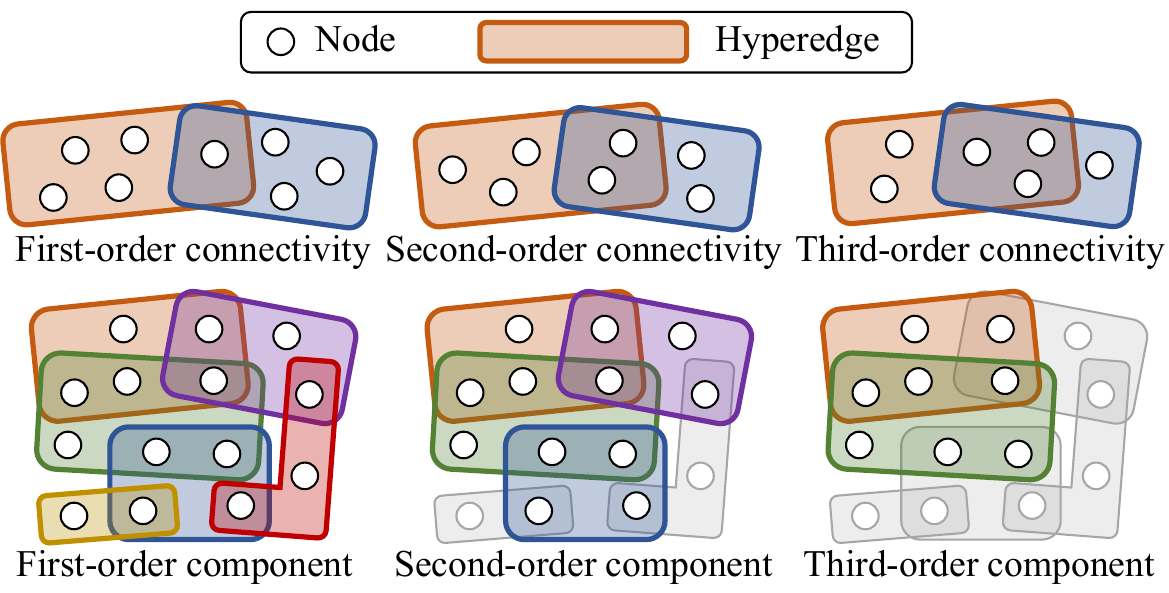}
\vspace*{-0.3cm}
\caption{Schematic illustration of the \mbox{$m$-th-order} connectivity and the \mbox{$m$-th-order} component.}
\label{Fig:Definition}
\end{figure}

As in networks, the existence of the giant component is a minimum condition for the collective functioning of hypergraphs.
In hypergraph, however, the notion of a connected component acquires an important new dimension.
In a network, only one node can be shared between two edges, whereas in a hypergraph, an arbitrary number of nodes can be shared between hyperedges.
The number of common nodes has a physical meaning which is the degree of cooperativity between hyperedges.
Therefore, neglecting them is to ignore an essential characteristic of hypergraphs.
To this end, we introduce the \mbox{$m$-th-order} connectivity in hypergraphs as the connectivity between two hyperedges sharing $m$ common nodes and \mbox{$m$-th-order} component as the connected component only through \mbox{$m$-th-} or higher-order connectivities as shown in Fig.~\ref{Fig:Definition}.
We call the components with $m \geq 2$ to be the higher-order components (HOCs)~\cite{2018CooleyThe, 2020AksoyHypernetwork}.

\begin{figure*}
\centering
\includegraphics[width=0.96\textwidth]{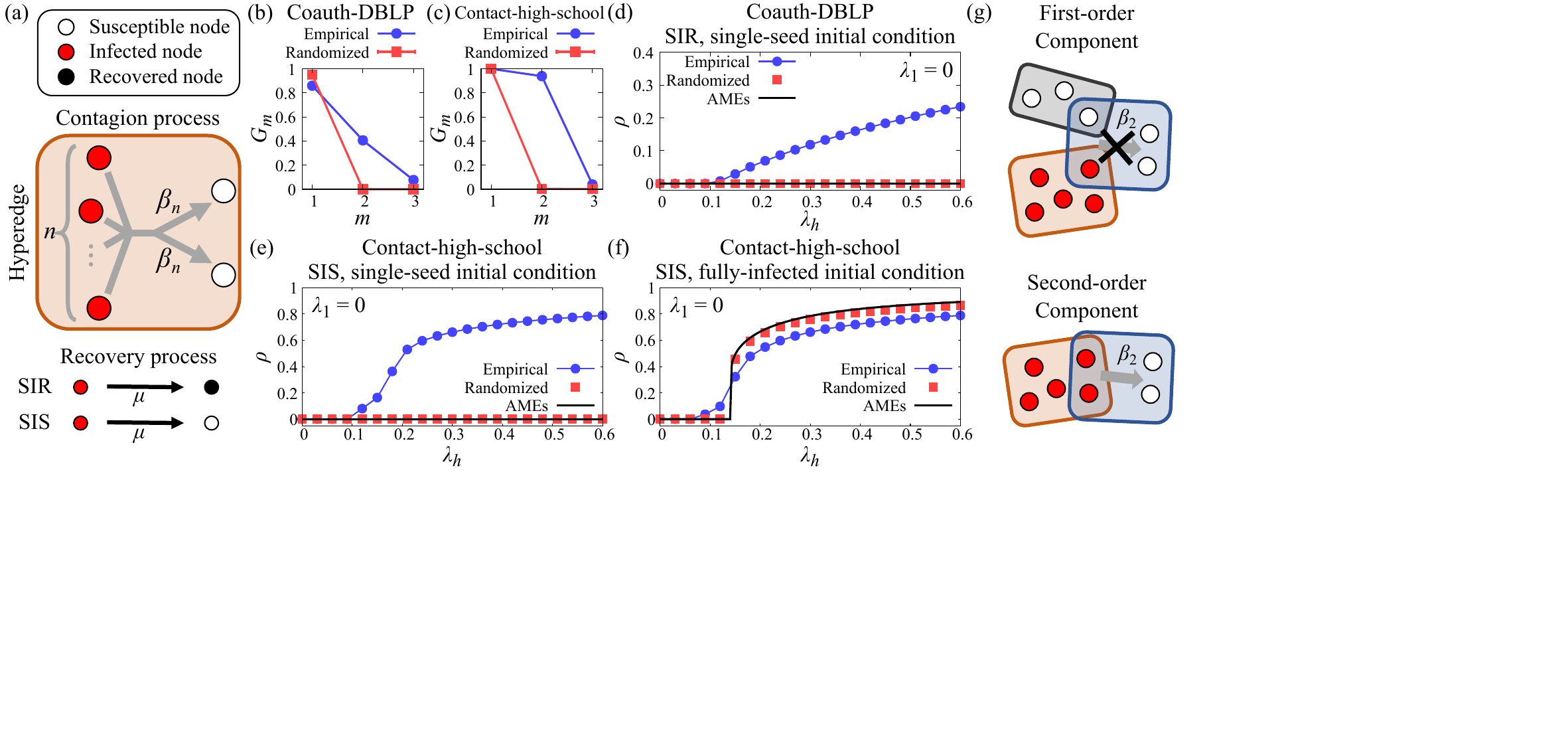}
\vspace*{-0.2cm}
\caption{
(a) Schematic illustration of the higher-order contagion process in a hyperedge.
(b-c) The relative size of the largest \mbox{$m$-th-order} component ($G_{m}$) on the empirical and randomized Coauth-DBLP hypergraph (the number of node $N=1,930,378$, the number of hyperedge $H=2,467,396$, mean degree $\left< k \right>=4.01$, and mean size $\left< s \right>=3.14$) and Contact-high-school hypergraph ($N=327$, $H=7,818$, $\left< k \right>=55.63$, $\left< s \right>=2.33$).
(d) The relative outbreak size ($\rho$) of higher-order SIR dynamics on Coauth-DBLP hypergraph with the single-seed initial condition.
(e) The relative outbreak size of higher-order SIS dynamics on Contact-high-school hypergraph with the single-seed initial condition, and (f) with the fully-infected initial condition.
We ran over minimum $10^{2}$ to maximum $10^{7}$ realizations and averaged over samples only with a relative outbreak size $\rho>10^{-3}$ for SIR dynamics and only with active phase samples with minimum $10^{2}$ to maximum $10^{3}$ steps after the relaxation steps minimum $10^{3}$ to maximum $10^{4}$ for SIS dynamics.
(g) A schematic illustration of the higher-order contagion process on the first- and second-order components.}
\vspace*{-0.3cm}
\label{Fig:RealWorldHypergaphs}
\end{figure*}

How the network structure affects its function is a longstanding problem in network science.
In this Letter, we demonstrate that the giant HOC serves as a structural backbone of higher-order contagion dynamics, a crucial example of collective behaviors.
First, we apply the higher-order contagion dynamics on two real-world hypergraphs and their randomized counterparts to reveal the effects of the giant HOC on higher-order contagion dynamics, and we confirm that the giant HOC functions as the channel of higher-order contagions from a single hyperedge infection source.
To verify whether this phenomenon is genuinely caused by HOC or due to other real-world hypergraph properties, we propose a novel random hypergraph model in which tunable and analytically calculable giant HOC exists.
We use this model to reveal that the giant HOC genuinely dictates the higher-order contagion dynamics.
We also confirm that the giant HOC ubiquitously exists in real-world hypergraphs from various fields, showing that the effect of the giant HOC structure on hypergraphs' function is a practical problem. 
This suggests that understanding the effects of the giant HOC on other diverse higher-order dynamics will be a fundamental problem in network science.

\textit{Higher-order contagion dynamics.}---We applied the higher-order contagion dynamics as an archetypical example of spreading phenomena~[Fig.~\ref{Fig:RealWorldHypergaphs}(a)].
In the higher-order contagion dynamics, the infection rate $\beta_{n}$ is determined according to the number of infected nodes $n$ in a hyperedge \cite{2022St-OngeInfluential}, and with this infection rate, each susceptible node in the hyperedge becomes infected independently.
With rate $\mu$, each infected node either becomes recovered and no longer participates in the contagion dynamics in susceptible-infected-recovered (SIR) dynamics or becomes susceptible in susceptible-infected-susceptible (SIS) dynamics.
We define the rescaled infection rate $\lambda_{n}=\beta_{n}/\mu$.
We call the infection caused by two or more infected nodes the higher-order contagion, and we set all $\beta_{n}$ of $n \geq 2$ to be the same $\beta_{h}$ for simplicity.
In this Letter, to understand the structural role of the giant HOC on the higher-order contagion dynamics, first, we focus on the higher-order contagion dynamics only ($\beta_{1}=0$, $\beta_{h} \geq 0$), then we investigate the more general case ($\beta_{1} \geq 0$, $\beta_{h} \geq 0$).

We analyze the higher-order contagion dynamics with Monte Carlo simulation and approximate master equations (AMEs)~\cite{2021St-OngeSocial, 2024St-OngeNonlinear}.
For the Monte Carlo simulations, we use two initial conditions.
The first is what we call the single-seed initial condition, where all nodes are susceptible except the entire nodes within one randomly chosen hyperedge, which are initially infected.
The second is the fully-infected initial condition, where all nodes are initially infected.
At each step, we traverse all the hyperedges where infected nodes exist to determine whether to infect each susceptible node within them in the next step; next, we traverse all the infected nodes to determine whether to be recovered in SIR dynamics or to be susceptible in SIS dynamics in the next step; then update the state of all nodes at once.
For SIR dynamics, we averaged only over the samples with a relative outbreak size $\rho$ larger than a certain threshold; and for SIS dynamics, we averaged over samples not in the absorbing state after the relaxation.
$\mu$ is set to 0.05 per step for both SIR and SIS dynamics.
We applied the AMEs following~\cite{2021St-OngeSocial, 2024St-OngeNonlinear} to obtain analytic results for dynamics on randomized hypergraphs.

\textit{Real-world hypergraphs.}---We applied the higher-order SIR dynamics on the Coauth-DBLP hypergraph~\cite{2018BensonSimplicial}, in which nodes represent authors and hyperedges represent publications recorded on DBLP and applied the higher-order SIS dynamics on the Contact-high-school hypergraph~\cite{2018BensonSimplicial}, in which nodes are highschoolers and hyperedges are maximal proximity groups during twenty seconds intervals.
We preprocessed the data by reducing the duplicated hyperedges with the same nodes set into a unique hyperedge, and we refer to such preprocessed hypergraphs as the empirical hypergraphs.
In these two empirical hypergraphs, there exists the extensive largest HOC [Fig.~\ref{Fig:RealWorldHypergaphs}(b, c)].
Randomized hypergraphs with the same degree and size distribution are used as null models, which do not contain the extensive largest HOC.
When constructing the randomized surrogates following the method proposed in Ref.~\cite{2022St-OngeInfluential}, the Coauth-DBLP hypergraph was randomized with the preserved number of nodes and hyperedges, and the Contact-high-school hypergraph, due to its small size, was randomized after expanding the number of nodes and hyperedges by ten times to suppress the largest HOC [Fig.~\ref{Fig:RealWorldHypergaphs}(b, c)].

The results of the higher-order SIR dynamics with the single-seed initial condition and $\lambda_{1}=0$ on the Coauth-DBLP hypergraph are shown in Fig~\ref{Fig:RealWorldHypergaphs}(d).
Higher-order contagion spreads above the finite critical $\lambda_{h}$ on the empirical hypergraph, which has an extensive HOC.
On the other hand, on randomized hypergraph, which does not have an extensive HOC, higher-order contagion could not spread even with much greater $\lambda_{h}$.

The results of the higher-order SIS dynamics with the single-seed initial condition and $\lambda_{1}=0$ on the Contact-high-school hypergraphs are qualitatively similar [Fig~\ref{Fig:RealWorldHypergaphs}(e)].
As in SIR dynamics, with extensive HOC, higher-order contagion spreads above the finite critical $\lambda_{h}$, but without extensive HOC, higher-order contagion could not spread from the single-seed initial condition.
On the contrary, SIS dynamics with the fully-infected initial condition exhibit different stationary states [Fig~\ref{Fig:RealWorldHypergaphs}(f)].
In this case, even if the extensive HOC is absent, higher-order contagion has a finite critical $\lambda_{h}$.
Therefore, a bistable region appears in the randomized hypergraph.
Note that there is also a small bistable region in the empirical hypergraph.

In sum, the extensive HOC is required to spread higher-order contagion from a single seed.
As shown in Fig.~\ref{Fig:RealWorldHypergaphs}(g), if only a first-order component exists, two independent infection routes are needed for higher-order contagion, which is improbable with a single-seed initial condition.
However, if there is a HOC, only one infection route can cause higher-order contagion.
Thus, higher-order contagion can spread from a single-seed initial condition.
On the other hand, if an infection is already prevalent, it is possible to maintain infection with only higher-order contagion since two infection routes can be secured even with only the first-order component.

Finally, we investigate the seventeen real-world hypergraph data from Ref.~\cite{2018BensonSimplicial} to identify if an extensive HOC is present in the other real-world hypergraphs.
These data were collected from various fields such as coauthorship, social contact, email, and online posts.
The number of nodes $N$ in each dataset ranges from 143 to 2,675,969, and the number of hyperedges $H$ ranges from 1,090 to 9,705,575.
We confirmed that the relative size of the largest second-order component is greater than 0.1 in fifteen out of seventeen empirical hypergraphs.
Therefore, it is practically important that the HOC is the backbone of higher-order contagion dynamics.
A detailed description of data and results are provided in Supplementary Material~\cite{SM}.

\begin{figure}
\centering
\includegraphics[width=0.48\textwidth]{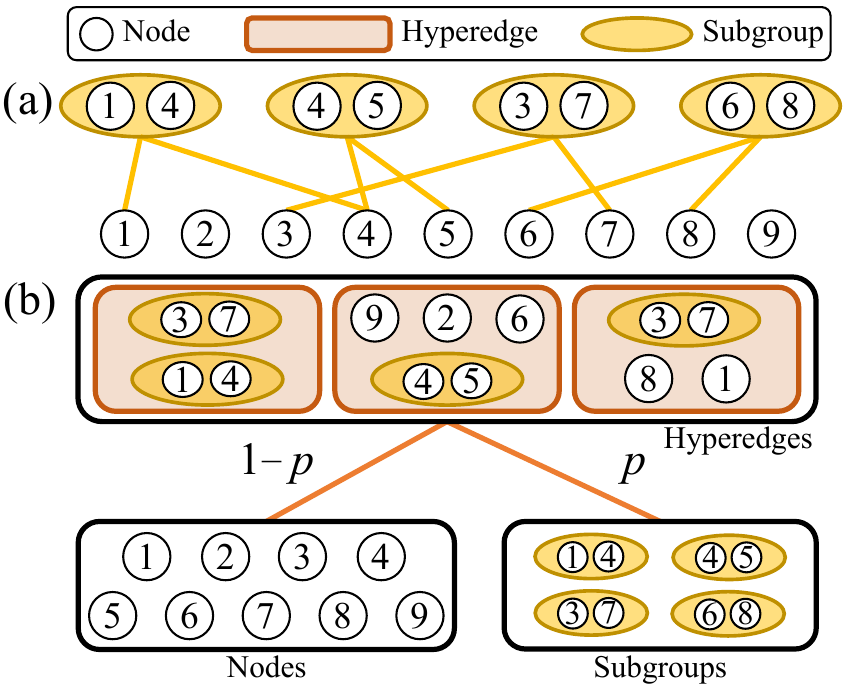}
\vspace*{-0.2cm}
\caption{
Illustration of the higher-order-connected hypergraph model for $N=9$, $H=3$, and $S=4$.
(a) Initially preassign $N=9$ nodes to $S=4$ subgroups.
For clarity, the number of nodes preassigned to each subgroup is fixed to two in this Letter.
(b) Each step, select a random node with probability $(1-p)$ or a random subgroup with probability $p$, and add it to a random hyperedge.}
\vspace*{-0.3cm}
\label{Fig:Higher-order-connectedHypergraphModel}
\end{figure}

\begin{figure*}
\centering
\includegraphics[width=0.96\textwidth]{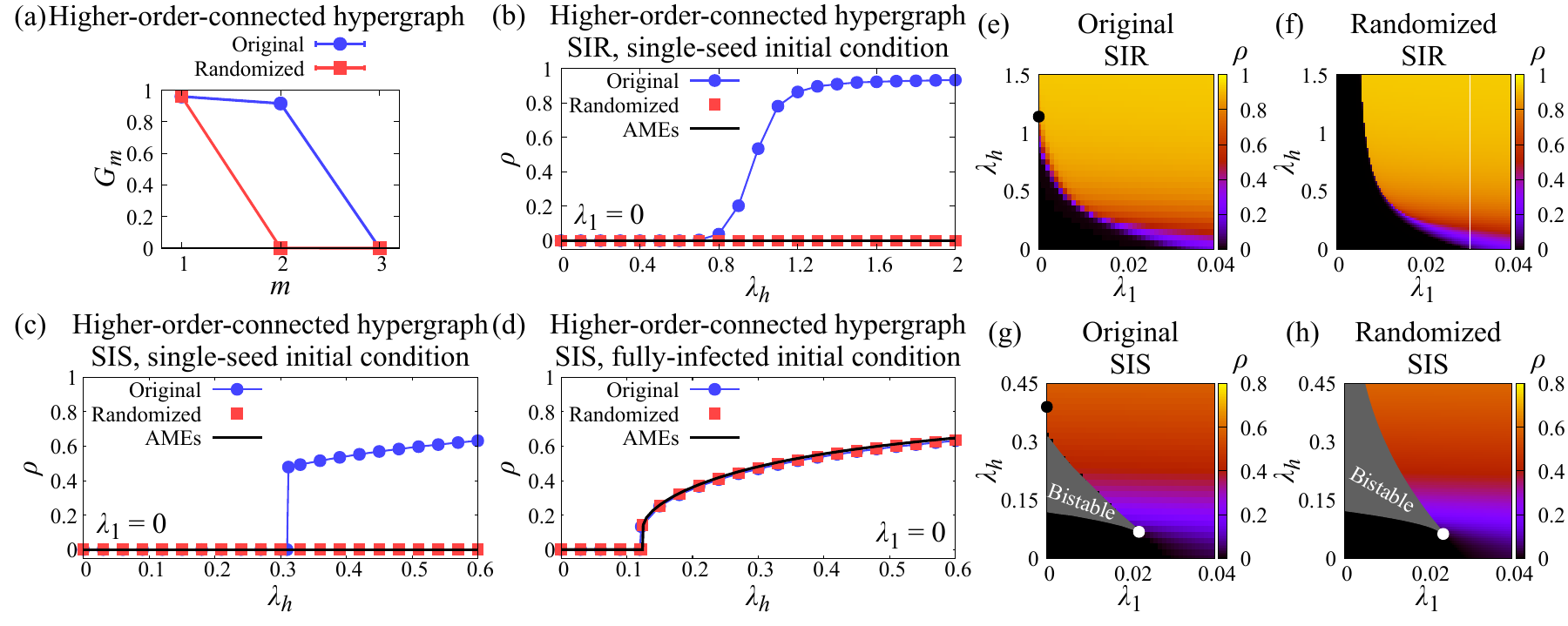}
\vspace*{-0.2cm}
\caption{
(a) The relative size of the \mbox{$m$-th-order} component $G_{m}$ in the original and randomized higher-order-connected hypergraph ($N=10^{5}$, $H=10^{5}$, $S=10^{5}$, $\left< k \right>=5$, $\left< s \right>=5$, $p=0.5$).
(b-d) The relative outbreak size ($\rho$) of (b) the higher-order SIR dynamics with the single-seed initial condition, (c) the higher-order SIS dynamics with the single-seed initial condition, (d) the higher-order SIS dynamics with the fully-infected initial condition on the higher-order-connected hypergraph.
(e-h) The phase diagram of the higher-order SIR and SIS dynamics on the original and randomized higher-order connected hypergraph with the single-seed and the fully-infected initial conditions.
White circles indicate tricritical points, and black circles indicate critical points $\lambda^{c}_{h}$ in the thermodynamic limit.
We ran over minimum 90 to maximum $10^{7}$ realizations and averaged over samples only with a relative outbreak size $\rho>1/\sqrt{N}$ for the higher-order SIR dynamics and only with active phase samples with $0.1\times N$ steps after the relaxation steps $N$ for the higher-order SIS dynamics.}
\vspace*{-0.3cm}
\label{Fig:ModelHypergraphs}
\end{figure*}

\textit{Higher-order-connected hypergraph model.}---Many other structures, such as short loop, clustering, and assortativity, are ampliated in real-world hypergraphs.
To single out the giant HOC's effects on the higher-order contagion dynamics, a random hypergraph free from such confounding structures while keeping HOC is useful.
This goal cannot be reached using existing random uniform hypergraphs because it has been proven that the giant HOC does not exist in a random uniform hypergraph with finite mean degree $\left< k \right> \sim \mathcal{O}(1)$ in the thermodynamic limit~\cite{2018CooleyThe}.
To overcome this problem, we propose a novel random hypergraph model admitting giant HOC.

In this model, in addition to nodes, the subgroups to which nodes are preassigned are introduced.
We assume that nodes included in a subgroup have a close relationship.
For example, in the coauthorship hypergraph, a subgroup consists of colleagues who write many papers jointly.
In the social contact hypergraph, a subgroup may represent friends who meet frequently, and a subgroup in the hypergraph of tags on online posts compose of close topics.
There is a greater chance that the nodes belonging to a subgroup join in a hyperedge simultaneously.

Our model hypergraph evolves through the following process.
First, prepare $N$ nodes, $H$ hyperedges, and $S$ subgroups.
We will present the results of $H/N=1$ and $S/N=1$.
We confirmed that qualitatively similar results are obtained using different values.
Second, randomly chosen nodes are preassigned to each subgroup according to the subgroup's size without duplication [Fig.~\ref{Fig:Higher-order-connectedHypergraphModel}(a)].
Here, the size of subgroups can have an arbitrary distribution, but in this Letter, the size of subgroups is fixed to two for clarity.
Finally, in the assignment process, the hypergraph evolves by recruiting either a random node (with probability $1-p$) or a random subgroup (with probability $p$) to a random hyperedge until the desired mean degree $\left< k \right>$ is reached [Fig.~\ref{Fig:Higher-order-connectedHypergraphModel}(b)].
Note that with probability $1-p$, one node is assigned to the hyperedge, and with probability $p$, all the nodes in a subgroup are assigned to the hyperedge in each assignment process.
The assignment is rejected if a newly recruited node already exists on the hyperedge.
When $p=0$, this evolving process is similar to the process of making the ER-like bipartite network~\cite{1960ErdosOn}.
As the model parameter $p$ increases, the probability that subgroups are selected increases, and thus the second-order connectivities are encouraged more.
Self-consistent equations to calculate the giant HOC size and equations for degree and size distribution of the model are provided in Supplementary Material~\cite{SM}.

We applied the higher-order contagion dynamics on the higher-order-connected hypergraphs and obtained consistent results with real-world hypergraphs.
The original higher-order-connected hypergraphs have the giant HOC, and the randomized counterparts have no giant HOC [Fig.~\ref{Fig:ModelHypergraphs}(a)].
The giant HOC was required for the higher-order contagion to spread from a single seed; and the giant HOC was not essential for infection to remain in the presence of many infected nodes [Fig.~\ref{Fig:ModelHypergraphs}(b-d)].
This suggests that the presence or absence of the giant HOC is the main determinant for the results we checked in the real-world hypergraph.

Finally, we investigated the more general case of combining simple contagion caused by one infected node and higher-order contagion, and we found fundamental differences between original and randomized higher-order-connected hypergraphs in phase diagrams [Fig.~\ref{Fig:ModelHypergraphs}(e-h)].
The phase diagrams of the original higher-order-connected hypergraph are the results of Monte Carlo simulation with $N=10^5$, and the phase diagrams of the randomized hypergraph are the results of AMEs.
When the giant HOC exists [Fig.~\ref{Fig:ModelHypergraphs}(e, g)], the infection can spread only by high-order contagion ($\beta_{1}=0$) with single-seed initial conditions, so the corresponding phase transition line touches the y-axis.
However, when the giant HOC does not exist [Fig.~\ref{Fig:ModelHypergraphs}(f, h)], the phase transition line of the single-seed initial condition never touches the y-axis because, in this case, the infection cannot spread only by higher-order contagion.

We found that there is a large finite-size effect on the critical point of higher-order contagion dynamics $\lambda^{c}_{h}$ in the case of the single-seed initial condition (so-called the invasion threshold) and performed finite-size scaling analysis with the following assumption,
\begin{equation}
\lambda^{c}_{h} - \lambda^{c}_{h}(N) \sim N^{-\theta},
\label{Eq:FSS}
\end{equation}
as provided in Supplementary Material~\cite{SM}.
We used $\lambda^{c}_{h}(N)$ as the point where the sample-to-sample fluctuation of the $\rho$ is the largest in the higher-order SIR model and as the smallest value at which the active phase sample was observed after the relaxation in the higher-order SIS model.
$\lambda^{c}_{h}$ obtained through finite-size scaling is marked by a black circle in the phase diagrams [Fig.~\ref{Fig:ModelHypergraphs}(e, g)].
Our results suggest that the HOC dictates higher-order contagion dynamics in the thermodynamic limit.

\textit{Conclusion.}---In this Letter, we focused on how the giant HOC in hypergraphs, which has been neglected so far, dictates the higher-order contagion dynamics.
The giant HOC makes the higher-order contagion spread from a single infected hyperedge seed by admitting the higher-order connected path between hyperedges.
The giant HOC is ubiquitous in the real-world hypergraph, on which diseases, knowledge, and opinions can spread.
However, existing random hypergraphs, generally used as null models for real-world hypergraphs, are incomplete for understanding higher-order contagion dynamics because of their lack of the giant HOC.
Therefore, it would be advantageous to use a model in which giant HOC exists, such as the higher-order-connected hypergraph model proposed in this paper, to analyze the higher-order contagion in the real world.

The contagion dynamics we studied in this Letter is chosen as a canonical example of the numerous collective dynamics.
Just as the giant HOC has played a significant role in the contagion dynamics, it will likely play a crucial role in other collective dynamics, such as synchronization dynamics~\cite{2021ChutaniHysteresis}, and in statistical validations \cite{2021MusciottoDetecting} on real-world hypergraphs.

Finally, it is noteworthy that the idea of overlap has been exploited for link predictions \cite{2018BensonSimplicial}. 
The overlapness proposed in~\cite{2021LeeHow} is also related, the measurements of which for the three hypergraphs used in this study are presented in the Supplementary Material~\cite{SM}.
Additional higher-order correlation features of potential relevance include the higher-order motif~\cite{2020LeeHypergraph, 2022LotitoHigher-order} and the hyperedge nestedness~\cite{2023KimContagion}.

\textit{Acknowledgement.}---This work was supported in part by the National Research Foundation of Korea (NRF) grant funded by the Korea government (MSIT) (No. NRF-2020R1A2C2003669).

\bibliography{Hypergraph}
\end{document}


\title{Supplementary material for ``Higher-order components dictate higher-order contagion dynamics in hypergraphs"}
\author{Jung-Ho Kim}
\affiliation{Department of Physics, Korea University, Seoul 02841, Korea}
\author{K.-I. Goh}
\email{kgoh@korea.ac.kr}
\affiliation{Department of Physics, Korea University, Seoul 02841, Korea}
\affiliation{Department of Mathematics, University of California Los Angeles, Los Angeles, CA 90095, USA}
\date{\today}
\maketitle

This supplementary material for ``Higher-order components dictate higher-order contagion dynamics in hypergraphs" is organized as follows.
First, we describe the seventeen real-world hypergraph data used in the main paper in Sec.~\ref{Sec:RealWorldHypergraph}.
Next, we demonstrate the self-consistent equations for the giant first- and second-order component in Sec.~\ref{Sec:ModelComponent}, and finite-size scaling analysis results on Sec.~\ref{Sec:FSS}.
Finally, the overlapness of the hypergraphs used in the main paper is presented in Sec.~\ref{Sec:Overlapness}.

\begin{figure}
\centering
\vspace*{-0.2cm}
\includegraphics[width=0.84\textwidth]{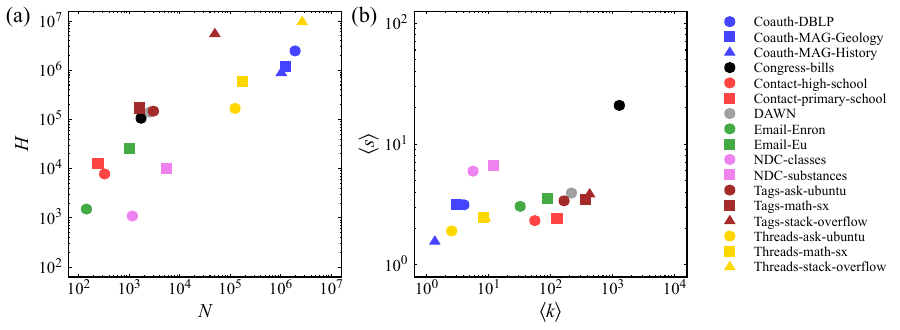}
\vspace*{-0.3cm}
\caption{
The basic statistics of the seventeen real-world hypergraphs without duplication of hyperedges.
(a) The number of nodes $N$ and the number of hyperedges $H$.
(b) The mean degree $\left< k \right>$ and the mean size $\left< s \right>$.
}
\label{Fig:BasicStatistics}
\end{figure}

\begin{figure}
\centering
\includegraphics[width=0.84\textwidth]{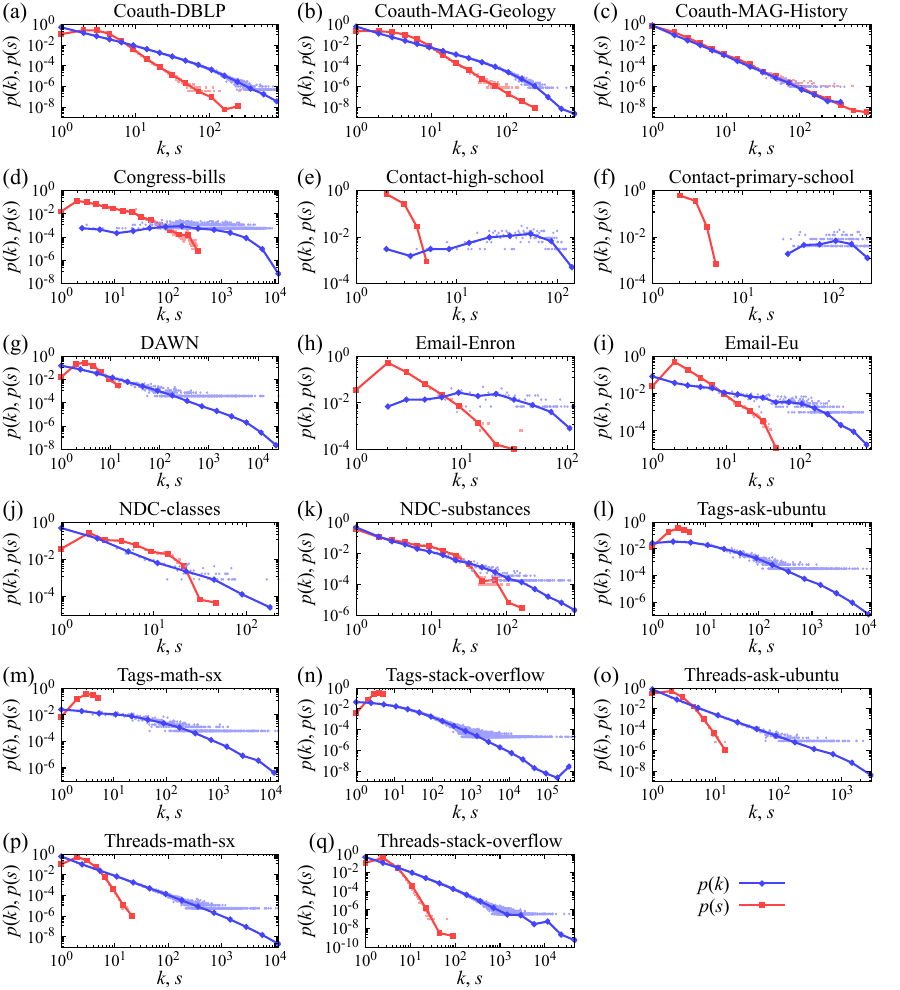}
\vspace*{-0.3cm}
\caption{The degree distribution $p(k)$ and the size distribution $p(s)$ of the seventeen real-world hypergraphs without duplication of hyperedges.}
\vspace*{-0.2cm}
\label{Fig:Distribution}
\end{figure}

\begin{figure}
\centering
\includegraphics[width=0.84\textwidth]{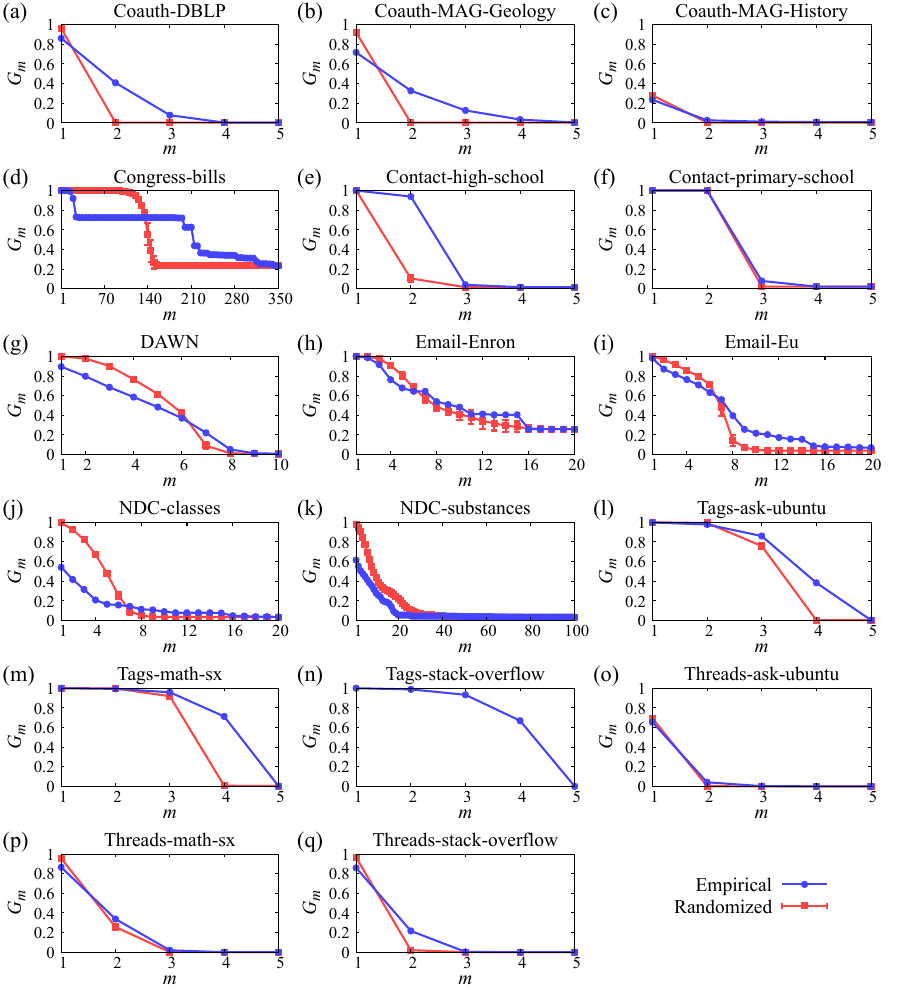}
\vspace*{-0.3cm}
\caption{
The relative size of the largest \mbox{$m$-th-order} component on empirical real-world hypergraph $G^{\text{empi}}_{m}$ and degree-and-size-preserved randomized hypergraph $G^{\text{rand}}_{m}$ of the seventeen real-world hypergraphs.
We also preserved the number of nodes and hyperedges during randomization processes.
We averaged the results of $10^{2}$ randomized hypergraphs, and the error bars represent the standard deviation.
We could not plot the data for the randomized Tags-stack-overflow hypergraph because the computational cost was too high.
}
\vspace*{-0.2cm}
\label{Fig:HOCProfile}
\end{figure}

\section{Real-world hypergraph data}
\label{Sec:RealWorldHypergraph}
We use the seventeen real-world hypergraph data from~\cite{2018BensonSimplicial}.
Detailed explanations for the real-world hypergraphs are in~\cite{dataset}.
However, we brought some descriptions for the sake of completeness of this Letter. 

\begin{itemize}
\item \textbf{Coauth-DBLP:} Nodes are authors and a hyperedge is a publication recorded on DBLP.

\item \textbf{Coauth-MAG-Geology~\cite{Sinha-2015-MAG}:} Nodes are authors and a hyperedge is a publication marked with the ``Geology" tag in the Microsoft Academic Graph.

\item \textbf{Coauth-MAG-History~\cite{Sinha-2015-MAG}:} Nodes are authors and a hyperedge is a publication marked with the ``History" tag in the Microsoft Academic Graph. 

\item \textbf{Congress-bills~\cite{Fowler-2006-connecting, Fowler-2006-cosponsorship}:} Nodes are US Congresspersons and hyperedges are comprised of the sponsor and co-sponsors of legislative bills put forth in both the House of Representatives and the Senate.

\item \textbf{Contact-high-school~\cite{Mastrandrea-2015-contact}:} Nodes are the people and hyperedges are maximal cliques of interacting individuals from an interval.

\item \textbf{Contact-primary-school~\cite{Stehl-2011-contact}:} Nodes are the people and hyperedges are maximal cliques of interacting individuals from an interval.

\item \textbf{DAWN:} Hyperedges in this dataset are the drugs used by a patient (as reported by the patient) in an emergency department visit.

\item \textbf{Email-Enron:} Nodes are email addresses at Enron and a hyperedge is comprised of the sender and all recipients of the email.

\item \textbf{Email-Eu~\cite{Leskovec-2007-evolution, Yin-2017-local}:} Nodes are email addresses at a European research institution. Hyperedges consist of a sender and all receivers such that the email between the two has the same timestamp.

\item \textbf{NDC-classes:} Each hyperedge corresponds to a drug and the nodes are class labels applied to the drugs.

\item \textbf{NDC-substances:} Each hyperedge corresponds to an NDC code for a drug, and the nodes are substances that make up the drug.

\item \textbf{Tags-ask-ubuntu:} Nodes are tags and hyperedges are the sets of tags applied to questions on askubuntu.com.

\item \textbf{Tags-math-sx:} Nodes are tags and hyperedges are the sets of tags applied to questions on math.stackexchange.com.

\item \textbf{Tags-stack-overflow:} Nodes are tags and hyperedges are the sets of tags applied to questions on stackoverflow.com.

\item \textbf{Threads-ask-ubuntu} Nodes are users on askubuntu.com, and a hyperedge comes from users participating in a thread that lasts for at most 24 hours.

\item \textbf{Threads-math-sx:} Nodes are users on math.stackexchange.com, and a hyperedge comes from users participating in a thread that lasts for at most 24 hours.

\item \textbf{Threads-stack-overflow:} Nodes are users on stackoverflow.com, and a hyperedge comes from users participating in a thread that lasts for at most 24 hours.
\end{itemize}

\section{Giant first- and second-order component in higher-order-connected hypergraph}
\label{Sec:ModelComponent}

\begin{figure}[b]
\centering
\includegraphics[width=0.9\textwidth]{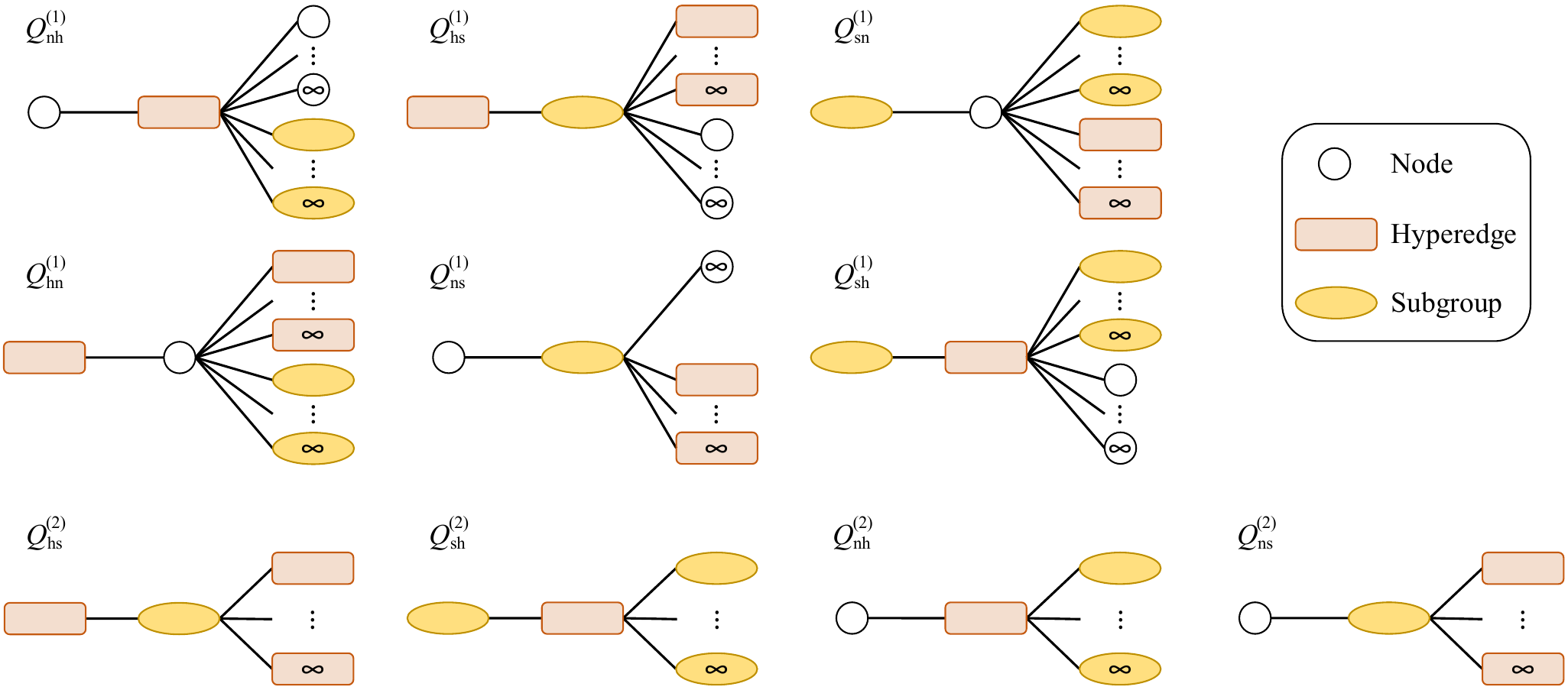}
\caption{A schematic illustration of Eqs.~(\ref{Eq:Q1s}, \ref{Eq:Q21}, \ref{Eq:Q22}).}
\vspace*{-0.2cm}
\label{Fig:Qs}
\end{figure}

The probability $G_{1}$ that a randomly chosen node belongs to the giant first-order component of the higher-order connected hypergraph can be calculated by the corresponding tripartite network consisting of nodes, hyperedges, and subgroups.
We hypothesized that the corresponding tripartite network is locally tree-like.
We introduce the probability $Q^{(1)}_{\text{xy}}$ that x is connected to the giant first-order component via y in the corresponding tripartite network [Fig.~\ref{Fig:Qs}], where $\text{x, y} \in \{\text{n, h, s}\}$ and n corresponds to node, h corresponds to hyperedge, and s corresponds to subgroup.
We can write the six self-consistent equations for the probabilities $Q^{(1)}_{\text{xy}}$ as follows,
\begin{equation}
\begin{aligned}
Q^{(1)}_{\text{nh}}&=\sum_{k_{\text{hn}}=0}^{\infty} \sum_{k_{\text{hs}}=0}^{\infty} \frac{k_{\text{hn}} p_{\text{hn}}(k_{\text{hn}})}{\left< k_{\text{hn}} \right>} p_{\text{hs}}(k_{\text{hs}}) [1-(1-Q^{(1)}_{\text{hn}})^{k_{\text{hn}}-1} (1-Q^{(1)}_{\text{hs}})^{k_{\text{hs}}}],\\
Q^{(1)}_{\text{hs}}&=\sum_{k_{\text{sh}}=0}^{\infty} \sum_{k_{\text{sn}}=0}^{\infty} \frac{k_{\text{sh}} p_{\text{sh}}(k_{\text{sh}})}{\left< k_{\text{sh}} \right>} p_{\text{sn}}(k_{\text{sn}}) [1-(1-Q^{(1)}_{\text{sh}})^{k_{\text{sh}}-1} (1-Q^{(1)}_{\text{sn}})^{k_{\text{sn}}}],\\
Q^{(1)}_{\text{sn}}&=\sum_{k_{\text{ns}}=0}^{\infty} \sum_{k_{\text{nh}}=0}^{\infty} \frac{k_{\text{ns}} p_{\text{ns}}(k_{\text{ns}})}{\left< k_{\text{ns}} \right>} p_{\text{nh}}(k_{\text{nh}}) [1-(1-Q^{(1)}_{\text{ns}})^{k_{\text{ns}}-1} (1-Q^{(1)}_{\text{nh}})^{k_{\text{nh}}}],\\
Q^{(1)}_{\text{hn}}&=\sum_{k_{\text{nh}}=0}^{\infty} \sum_{k_{\text{ns}}=0}^{\infty} \frac{k_{\text{nh}} p_{\text{nh}}(k_{\text{nh}})}{\left< k_{\text{nh}} \right>} p_{\text{ns}}(k_{\text{ns}}) [1-(1-Q^{(1)}_{\text{nh}})^{k_{\text{nh}}-1} (1-Q^{(1)}_{\text{ns}})^{k_{\text{ns}}}],\\
Q^{(1)}_{\text{ns}}&=\sum_{k_{\text{sn}}=0}^{\infty} \sum_{k_{\text{sh}}=1}^{\infty} \frac{k_{\text{sn}} p_{\text{sn}}(k_{\text{sn}})}{\left< k_{\text{sn}} \right>} p_{\text{sh}}(k_{\text{sh}}) [1-(1-Q^{(1)}_{\text{sn}})^{k_{\text{sn}}-1} (1-Q^{(1)}_{\text{sh}})^{k_{\text{sh}}}],\\
Q^{(1)}_{\text{sh}}&=\sum_{k_{\text{hs}}=0}^{\infty} \sum_{k_{\text{hn}}=0}^{\infty} \frac{k_{\text{hs}} p_{\text{hs}}(k_{\text{hs}})}{\left< k_{\text{hs}} \right>} p_{\text{hn}}(k_{\text{hn}}) [1-(1-Q^{(1)}_{\text{hs}})^{k_{\text{hs}}-1} (1-Q^{(1)}_{\text{hn}})^{k_{\text{hn}}}],
\label{Eq:Q1s}
\end{aligned}
\end{equation}
where $k_{\text{xy}}$ and $p_{\text{xy}}(k_{\text{xy}})$ are a x's network degree towards y in the corresponding tripartite network and its distribution.
$p_{\text{xy}}(k_{\text{xy}})$ are the Poisson distribution with $\left< k_{\text{nh}} \right> = \frac{1-p}{1+p} \left< k \right>$, $\left< k_{\text{hs}} \right> = \frac{N}{H} \frac{p}{1+p} \left< k \right>$, $\left< k_{\text{hn}} \right> = \frac{N}{H} \frac{1-p}{1+p} \left< k \right>$, $\left< k_{\text{ns}} \right> = \frac{2S}{N}$, and $\left< k_{\text{sh}} \right> = \frac{N}{S} \frac{p}{1+p} \left< k \right>$, and $p_{\text{sn}}(k_{\text{sn}})$ is a delta distribution with $p_{\text{sn}}(2)=1$.
Note that summation starts $k_{\text{sh}}=1$ in the equation for $Q^{(1)}_{\text{ns}}$.
This is because a subgroup only belongs to a component when included in at least one hyperedge.

Finally, the order parameter $G_{1}$ can be computed as follows,
\begin{equation}
G_{1}=1-\sum_{k_{\text{nh}}=0}^{\infty} \sum_{k_{\text{ns}}=0}^{\infty} [p_{\text{nh}}(k_{\text{nh}}) (1-Q^{(1)}_{\text{nh}})^{k_{\text{nh}}} p_{\text{ns}}(k_{\text{ns}}) (1-Q^{(1)}_{\text{ns}})^{k_{\text{ns}}}].
\label{Eq:G1}
\end{equation}

\begin{figure}[t]
\centering
\includegraphics[width=0.6\textwidth]{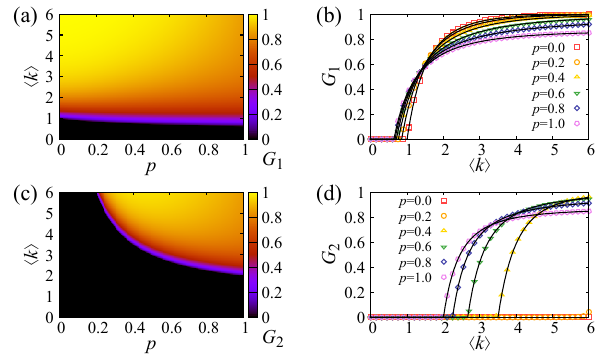}
\caption{
Plots of (a--b) $G_{1}$ and (c--d) $G_{2}$ of hypergraph with $H/N=1$ and $S/N=1$.
Heatmaps and lines are the results from Eq.~(\ref{Eq:G1}) and Eq.~(\ref{Eq:G2}), and symbols are Monte Carlo simulation results averaged of $10^{2}$ realizations with $N=10^{6}$.}
\vspace*{-0.2cm}
\label{Fig:GiantComponent}
\end{figure}

The probability $G_{2}$ that a randomly chosen node belongs to the giant second-order component of the higher-order connected hypergraph also can be calculated by the corresponding tripartite network consisting of nodes, hyperedges, and subgroups.
We hypothesized that the giant second-order component emerges only through subgroups, and the corresponding tripartite network is locally tree-like.
Corroboration by Monte Carlo simulations supports the validity of this assumption [Fig.~\ref{Fig:GiantComponent}(d)].
We can write self-consistent equations for the probability $Q^{(2)}_{\text{hs}}$ ($Q^{(2)}_{\text{sh}}$) that a hyperedge (subgroup) is connected to the giant second-order component via a subgroup (hyperedge) in the corresponding tripartite network [Fig.~\ref{Fig:Qs}] as follows,
\begin{equation}
\begin{aligned}
Q^{(2)}_{\text{hs}}&=\sum_{k_{\text{sh}}=0}^{\infty} \frac{k_{\text{sh}} p_{\text{sh}}(k_{\text{sh}})}{\left< k_{\text{sh}} \right>} [1-(1-Q^{(2)}_{\text{sh}})^{k_{\text{sh}}-1}],\\
Q^{(2)}_{\text{sh}}&=\sum_{k_{\text{hs}}=0}^{\infty} \frac{k_{\text{hs}} p_{\text{hs}}(k_{\text{hs}})}{\left< k_{\text{hs}} \right>} [1-(1-Q^{(2)}_{\text{hs}})^{k_{\text{hs}}-1}],
\label{Eq:Q21}
\end{aligned}
\end{equation}
where $k_{\text{hs}}$ ($k_{\text{sh}}$) and $p_{\text{hs}}(k_{\text{hs}})$ ($p_{\text{sh}}(k_{\text{sh}})$) are a hyperedge's (subgroup's) network degree towards subgroups (hyperedges) in the corresponding tripartite network and its distribution.

After solving for $Q^{(2)}_{\text{hs}}$ and $Q^{(2)}_{\text{sh}}$, one can calculate the probability $Q^{(2)}_{\text{nh}}$ ($Q^{(2)}_{\text{ns}}$)that a node is connected to the giant second-order component via a hyperedge (subgroup) in the corresponding tripartite network [Fig.~\ref{Fig:Qs}] as follows,
\begin{equation}
\begin{aligned}
Q^{(2)}_{\text{nh}}&=\sum_{k_{\text{hs}}=0}^{\infty} p_{\text{hs}}(k_{\text{hs}}) [1-(1-Q^{(2)}_{{\text{hs}}})^{k_{\text{hs}}}],\\
Q^{(2)}_{\text{ns}}&=\sum_{k_{\text{sh}}=0}^{\infty} p_{\text{sh}}(k_{\text{sh}}) [1-(1-Q^{(2)}_{{\text{sh}}})^{k_{\text{sh}}}].
\label{Eq:Q22}
\end{aligned}
\end{equation}

Finally, the $G_{2}$ can be computed as follows,
\begin{equation}
G_{2}=1-\sum_{k_{\text{nh}}=0}^{\infty} \sum_{k_{\text{ns}}=0}^{\infty} [p_{\text{nh}}(k_{\text{nh}}) (1-Q^{(2)}_{\text{nh}})^{k_{\text{nh}}} p_{\text{ns}}(k_{\text{ns}}) (1-Q^{(2)}_{\text{ns}})^{k_{\text{ns}}}],
\label{Eq:G2}
\end{equation}
where $k_{\text{nh}}$ ($k_{\text{ns}}$) and $p_{\text{nh}}(k_{\text{nh}})$ ($p_{\text{ns}}(k_{\text{ns}})$) are a node's network degree towards hyperedges (subgroups) in the corresponding tripartite network and its distribution.
The critical mean degree $\left< k \right> _{c}=\frac{\sqrt{HS}}{N}\frac{1+p}{p}$ at which the giant second-order components emerges can be calculated using Jacobian matrix of Eq.~(\ref{Eq:Q21}).

Plots of $G_{1}$ and $G_{2}$ are shown in Fig.~\ref{Fig:GiantComponent}(a--d).
The heatmap in (a) and (c), lines in (b) and (d) are the results from Eq.~(\ref{Eq:G1}) and Eq.~(\ref{Eq:G2}), and the symbols on (b) and (d) are the Monte Carlo simulation results averaged over $10^{2}$ realizations with $N=10^{6}$, and two results agree excellently well.
As shown in Fig.~\ref{Fig:GiantComponent}(a--d), $G_{2}$ can be controlled by adjusting the model parameter $p$.
In addition, as shown in (a) and (c), even for the systems with similar $G_{1}$, the value of $G_{2}$ can be significantly different.
Therefore, to fully understand the structure of the hypergraph, it is necessary to pay attention not only to the giant first-order component which has been studied routinely, but also to the giant HOCs that have been overlooked until now, and the proposed higher-order-connected hypergraph model provides a means to its systematic study.

\section{Finite-size scaling results}
\label{Sec:FSS}

\begin{figure}[h]
\centering
\includegraphics[width=0.6\textwidth]{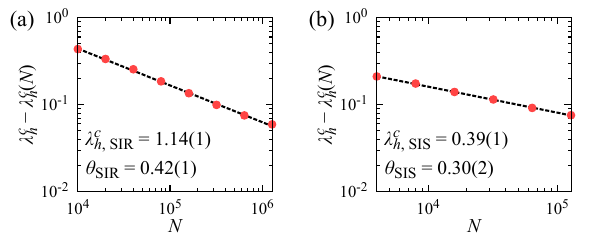}
\caption{
Finite-size scaling analysis results for $\lambda^{c}_{h}$ on the higher-order (a) SIR and (b) SIS dynamics.
Both panels represent the relation of Eq.~(1).
}
\vspace*{-0.2cm}
\label{Fig:FSS}
\end{figure}

Finite-size scaling analysis results for $\lambda^{c}_{h}$ on the higher-order SIR and SIS dynamics are shown in Fig.~\ref{Fig:FSS}.

\section{Overlapness}
\label{Sec:Overlapness}

\begin{figure}[h]
\centering
\includegraphics[width=0.9\textwidth]{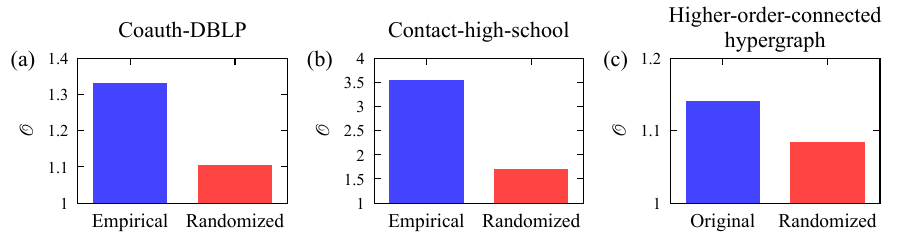}
\caption{
The overlapness of the hypergraphs used in the main paper.
The error bars represent the standard deviation.
}
\vspace*{-0.2cm}
\label{Fig:Overlapness}
\end{figure}

According to Ref.~[42], the overlapness $o$ of the hyperedge set $\mathcal{E}$ can be defined as follows,
\begin{equation}
o(\mathcal{E}) = \frac{\sum_{e \in \mathcal{E}} |e| }{| \cup_{e \in \mathcal{E}} e |}
\label{Eq:Overlapness}
\end{equation}
where the numerator denotes the sum of the size of hyperedges in the hyperedge set and the denominator indicates the number of unique nodes in the hyperedges belong to hyperedge set.
For example, for $\mathcal{E}=\{\{a, b, c\}, \{b, c, d, e, f\}\}$, $o({\mathcal{E}})=8/6$.
If there is no overlap in $\mathcal{E}$, $o(\mathcal{E})$ is 1, and as overlap increases, $o(\mathcal{E})$ increases.

We defined the overlapness of the hypergraph $\mathcal{O}$ as the average of the overlapness of the nodes in the hypergraph, and we defined the overlapness of the node as the overlapness of the set of hyperedges containing the node.
The overlapness of the hypergraphs used in the main paper are shown in Fig.~\ref{Fig:Overlapness}, and the randomized hypergraphs have lower overlapness than empirical or original ones.

\bibliography{Hypergraph_SM}